\documentclass[aoas]{imsart}
\usepackage{caption}
\usepackage{subcaption}
\RequirePackage{amsthm,amsmath,amsfonts,amssymb}
\RequirePackage[authoryear]{natbib}
\RequirePackage[colorlinks,citecolor=blue,urlcolor=blue]{hyperref}
\RequirePackage{graphicx}
\RequirePackage{booktabs}
\startlocaldefs

\endlocaldefs

\begin{document}

\begin{frontmatter}
\title{Latent Subgroup Identification in Image-on-scalar Regression}

\begin{aug}
\author[A]{\fnms{Zikai} \snm{Lin}\ead[label=e1]{zikai@umich.edu}},
\author[B]{\fnms{Yajuan} \snm{Si}\ead[label=e2,mark]{yajuan@umich.edu}}
\and
\author[A]{\fnms{Jian} \snm{Kang}\ead[label=e3,mark]{jiankang@umich.edu}}
\address[A]{Department of Biostatistics, University of Michigan, Ann Arbor, Michigan, U.S.A. 
\printead{e1,e3}}

\address[B]{Survey Research Center, University of Michigan, Ann Arbor, Michigan, U.S.A.
\printead{e2}}
\end{aug}

\begin{abstract}
Image-on-scalar regression has been a popular approach to modeling the association between brain activities and scalar characteristics in neuroimaging research. The associations could be heterogeneous across individuals in the population, as indicated by recent large-scale neuroimaging studies, e.g., the Adolescent Brain Cognitive Development (ABCD) study. The ABCD data can inform our understanding of heterogeneous associations and how to leverage the heterogeneity and tailor interventions to increase the number of youths who benefit. It is of great interest to identify subgroups of individuals from the population such that: 1) within each subgroup the brain activities have homogeneous associations with the clinical measures; 2) across subgroups the associations are heterogeneous; and 3) the group allocation depends on individual characteristics. Existing image-on-scalar regression methods and clustering methods cannot directly achieve this goal. We propose a latent subgroup image-on-scalar regression model (LASIR) to analyze large-scale, multi-site neuroimaging data with diverse sociodemographics. LASIR introduces the latent subgroup for each individual and group-specific, spatially varying effects, with an efficient stochastic expectation maximization algorithm for inferences. We demonstrate that LASIR outperforms existing alternatives for subgroup identification of brain activation patterns with functional magnetic resonance imaging data via comprehensive simulations and applications to the ABCD study. We have released our reproducible codes for public use with the software package available on \href{https://github.com/zikaiLin/lasir}{Github}.
\end{abstract}


\end{frontmatter}


\section{Introduction}
\label{s:intro}

The rapid development of human neuroimaging techniques and analytic methods has provided unprecedented information to social and health sciences research. For example, the Adolescent Brain Cognitive Development (ABCD) study has collected a broad range of imaging outcomes of 11,875 U.S. children aged 9/10 years old for child development study \citep{casey2018adolescent}. The collected functional magnetic resonance imaging (fMRI) data allow us to examine the association between cognitive performances and brain activities with the blood-oxygen-level-dependent (BOLD) signal at voxels.

With brain imaging intensity outcomes and scalar covariates, the mass univariate analysis (MUA) fits regression models independently on every voxel, accounts for spatial dependence in test statistics, and applies multiple testing adjustment methods such as family-wise error rate or false discovery rate (FDR) to select voxels with significant signals~\citep{li2011multiscale, jenkinson2012fsl, groppe2011mua}. MUA has been widely used and implemented in software, e.g., the FMRIB Software Library and the Statistical Parametric Mapping software~\citep{jenkinson2012fsl, ashburner2012spm}. Alternative to MUA methods, image-on-scalar regression models image outcomes as multidimensional arrays or tensors and scalar predictors and preserves the inherited adjacency structure of the voxels \citep{li2020sparse, zhang2020nnisr, liu2017tensor-on-scalar}. However, we are faced with several challenges when applying existing image-on-scalar regression models to large-scale heterogeneous imaging data.

First, the patterns of brain signals are heterogeneous across subgroups of individuals. The imaging research literature includes some existing clustering or subgroup identification methods. Independent component analysis attempts to cluster similar features within the fMRI data, but not the individuals \citep{mckeown2003ICA}. \citet{lecei2019subgroupIdentificationNeuroimaging} have applied latent class analysis of brain activities to identify subgroups of children with Attention-deficit/hyperactivity disorder symptoms. \citet{brodersen2011generative} have used a combination of dynamic causal models and support vector machines for a subject-wise fMRI data classification. 
Nevertheless, the limitation of machine learning methods lies in difficulties in model interpretation, the adjustment of potential associations between scalar predictors, and the capability to make statistical inferences.

Second, associations between brain activities and clinical/sociodemographic information may present individual-level heterogeneity. Most imaging clustering approaches focus on the segmentation and intensity of images but rarely pay attention to the heterogeneous association between brain images and scalar predictors. The sociodemographic, environmental, or epidemiological factors could shape the function and structure of the human brain and modify the associations between brain activities and the exposure variable of interest \citep{paus2010populationNeuroscience, zuo2018developmental, falk2013representative}. \citet{assari2021parental} have identified a positive association between parental educational attainment and children's superior temporal cortical surface area. In their analysis, however, the cortical surface area is considered a continuous outcome variable, where limited information is extracted from the brain imaging data. Moreover, reproducible brain-wide association studies require large sample sizes, especially to account for the subject-level heterogeneity~\citep{Marek2022}. 

In addition, the spatial dependence of image intensity among voxels is complex and heterogeneous. Failing to account for such spatial dependence might compromise the power of statistical findings. \citet{zhu2014spatially} have developed a spatially varying coefficient model (SVCM) to incorporate both spatial dependence and piece-wise smooth covariate effects. Improving SVCM with sparsity and piece-wise smoothness, \citet{shi2015thresholded} use thresholded multiscale Gaussian processes as non-parametric priors imposed on the coefficient functions. Furthermore, \citet{zhang2020nnisr} have built neural networks to model the complex spatial dependence. However, none of these methods focus on subgroup detection. 

Targeting the U.S. population, the large-scale ABCD study has collected brain images and biospecimens of 11,875 children aged 9-10 from 21 sites across the U.S. for environmental exposure, neuroimaging, and substance use analysis. In this paper, we aim to study the associations between voxel-level brain activities and the general cognitive ability measure, e.g., the $g$-factor \citep{akshoomoff2013cognitive}, which is the exposure variable of interest. Sociodemographics are expected to moderate the relationship between cognitive ability and neural measures in image data modeling.  We propose a novel LAtent Subgroup identification in Image-on-scalar Regression (LASIR) model to account for the heterogeneity in both individual brain imaging intensities and their associations with the exposure variable of interest, in the adjustment of other individual-level characteristics. Our setting can be extended to model heterogeneous associations between brain imaging outcomes and multiple variables of interest. As an illustration, our analysis focuses on the exposure variable of researchers' interest in the ABCD study is the $g$-factor. We assume that 1) all individuals are assigned into a finite number of latent groups; 2) the brain imaging intensities and the spatially varying coefficients (SVC) of the exposure variable of interest are group-specific; and 3) the allocation probabilities of subgroups depend on the adjustment variables. We adopt the basis function expansion approach to the SVC modeling and develop a stochastic expectation maximization (SEM) algorithm for inferences \citep{diebolt1996stochastic}. 

The remainder of the manuscript is organized as follows. We introduce the LASIR model framework and provide details about parameter estimation using SEM and model selection criteria in Section \ref{s:model}. We compare the performance of LASIR  with alternative approaches via simulations in Section \ref{s:simulation} and analyses of the ABCD data in Section \ref{ss:rda}. Section \ref{s:discuss} concludes with discussions on directions of future work.

\section{Methods}
\label{s:model}
We begin with the basic notation.  Let $\mathbb{R}^d$ be a $d$-dimension vector space of real values for any positive integer $d$. Denote by $\mathcal{N}_d(\nu, \Sigma)$ a $d$-dimensional multivariate normal distribution with mean vector $\nu\in \mathbb{R}^d$ and the covariance matrix $\Sigma \in \mathbb{R}^{d\times d}$. Denote by $\mathcal{GP}(\mu(\cdot),\kappa(\cdot, \cdot))$ a Gaussian process (GP) with the mean function $\mu(\cdot):\mathbb{R}^d \rightarrow \mathbb{R}$ and covariance kernel $\kappa(\cdot, \cdot): \mathbb{R}^{d\times d} \rightarrow \mathbb{R}$. And let $f_d(\cdot): \mathbb{R}^d \rightarrow \mathbb{R}$ be the multivariate Gaussian density function.

\subsection{Latent Subgroup Identification in Image-on-scalar Regression}
\label{sec:lasir}
Suppose the data are collected from $n$ individuals in $S$ different study sites. For individual $i\in \{1, \dots, n\}$, let $\mathbf{u}_i\in \{0,1\}^S$ be a vector of binary site indicators for individual $i$, $u_{is} = 1$ if the $i$-th individual comes from site $s\in \{1,\ldots,S\}$, otherwise $u_{is} = 0$. Let $\mathcal{V}=\{v_m\in\mathbb{R}^3 \}_{m=1}^d$ 
represent a collection of $d$ voxels in the brain regions of interest.  Let $y_i(v) \in \mathbb{R}$ denote the image measurement at voxel $v\in \mathcal{V}$. Let $\mathbf{x}_i = (x_{ij})_{j=0}^{p}$ be a $(p+1)$-dimensional vector containing $p$ exposure variables of interest and intercept. And let $\mathbf{z}_i=(z_{ir})_{r=1}^q$ be a $q$-dimensional vector containing $q$ adjustment or control variables.   

Our proposed LASIR, as a latent subgroup identification in image-on-scalar regression model, assumes all individuals belong to a finite number of $K$ subgroups, where the subgroup indicator $\boldsymbol{\delta}_i = \left(\delta_{ik}\right)_{k=1}^K \in \{0,1\}^K$ of individual $i$ is denoted as a $K$-dimensional vector with $\delta_{ik} =1$ if individual $i$ belongs to subgroup $k$, otherwise $\delta_{ik} =0$. The multivariate brain imaging outcome intensity $y_i(v)$ of individual $i$ at voxel $v$ is composed of three different mean components: (1) Subgroup-specific effects, which includes subgroup-specific intercepts and coefficients of the exposure variables $\mathbf{x}_i$; (2) Fixed effects of the control variables $\mathbf{z}_i$; and (3) Site-specific fixed effects $\mathbf{u}_i$. The model can be generalized by replacing the fixed effects with random effects. 

Specifically, we represent our model as follows:
\begin{align}
\label{mod:1st}
         y_i(v) &= \sum_{j=0}^p x_{ij} \beta_{ij}(v) + \sum_{r = 1}^q z_{ir}\eta_r(v)+
        \sum_{s = 1}^Su_{is}\gamma_{s}(v) + \epsilon_i(v),\quad \epsilon_i(v)  \sim \mathcal{GP}(0, \kappa),\\
     \nonumber   \beta_{ij}(v) &= \sum_{k=1}^K\delta_{ik}\alpha_{kj}(v),\quad 
   \textrm{Pr}(\delta_{ik}=1) =  \dfrac{\exp(\mathbf{w}_k^\top \mathbf{z}_i)}{\sum_{c=1}^K \exp(\mathbf{w}_c^\top \mathbf{z}_i)},
  \end{align}
for $i = 1,\ldots, n$, $j = 0,\ldots p$, and $k=1, \ldots, K$. Here the SVCs  $\{\beta_{ij}(\cdot)\}_{j=0}^{p}$ are individual-specific, determined by the group indicator $\delta_{ik}$ and group-specific SVCs $\{\alpha_{kj}(\cdot)\}_{j=0}^{p}$. Meanwhile, $\gamma_s(\cdot)$ is a site-specific coefficient function, and $\eta_r(\cdot)$ is the coefficient function of control variable $z_{ir}$, both of which are considered as fixed effects and do not vary across subgroups. 
The latent subgroup indicator $\boldsymbol{\delta}_i$ is determined by a mapping function of control variables $\mathbf{z}_i$ and unknown coefficients or weights $\{\mathbf{w}_k\}_{k=1}^K$, as covariate-dependent latent profiling~\citep{missEHR-Si19,blpm-si:15}. Considering identification, we set $\mathbf{w}_K\equiv \vec{0}$.  To account for the spatial dependence of image outcomes, we assume that the random error is spatially correlated across voxels and follows a GP with mean zero and covariance kernel $\kappa$: $\epsilon_i(v)\sim \mathcal{GP}(\mathbf{0}, \kappa)$, where the details about the kernel $\kappa$ specification are given below in Section~\ref{assumption}. 

\begin{figure}
    \includegraphics[width=5in]{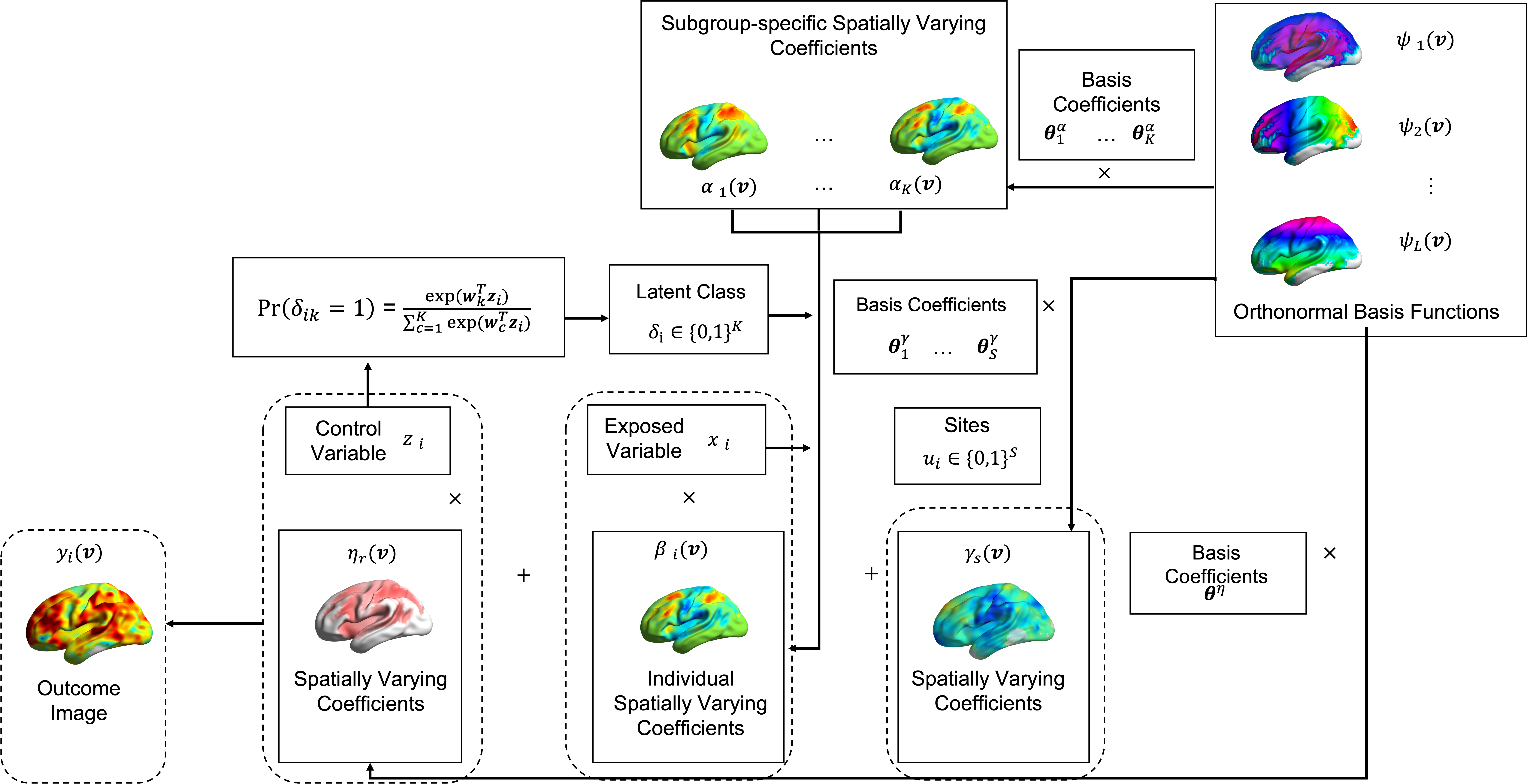}
\caption{Graphical representation of the LAtent Subgroup Image-on-Scalar Regression model (LASIR). Suppose there are $K$ subgroups in the population, and individual $i$ is assigned into subgroup $k$. The outcome image measure of individual $i$ at voxel $v$, denoted as $y_i(v)$, is composed of three mean components: 1) group-specific intercept $\alpha_{k0}(v)$ and group-specific effect $\left\{\alpha_{kj}(v)\right\}_{j=0}^p$ associated with the exposure variable $\mathbf{x}_i = \{x_{ij}\}_{j=0}^{p}$ of subgroup $k$; 2) fixed effect $\eta(v) = \{\eta_{r}(v)\}_{r=1}^q$ and associated control variables $\mathbf{z}_i = \{z_{ir}\}_{r=1}^q$; and 3) site-specific effect $\gamma(v) = \{\gamma_s(v)\}_{s=1}^S$ associated with site indicator $u_{is} \in \{0,1\}$, for $s = 1, \ldots, S$. The latent subgroup indicator $\delta_i \in \{0,1\}^K$ of individual $i$ is determined by a multinomial-logit model with coefficients $\mathbf{w} = \{\mathbf{w}_k\}_{k=1}^K$ and control variables $\mathbf{z}_i$. All three spatial varying coefficients: $\alpha_{kj}(v)$, $\gamma_s(v)$, and $\eta_r(v)$, are approximated by the basis function expansion with basis coefficients: $ \{\boldsymbol{\theta}_{k}^\alpha\}_{k=1}^K$, $\{\boldsymbol{\theta}_s^\gamma\}_{s=1}^S$, and $\{\boldsymbol{\theta}_r^\eta\}_{r=1}^q$, respectively, and a pre-computed set of $L$ constant basis functions $\{\psi_l(v)\}_{l=1}^L$, evaluated at voxel $v$.}
    \label{fig:model_illustration}
\end{figure}

Figure \ref{fig:model_illustration} illustrates the graphical representation of LASIR.

\subsection{Model Assumption}
\label{assumption}
We adopt the basis expansion approach to modeling SVCs in \eqref{mod:1st}. Consider $L$ pre-specified orthonormal basis functions $\{\psi_l(v)\}_{l=1}^L$ that satisfy  $\int_{\mathcal{V}}\psi^2_l(v) d v = 1$ for $l = 1,\ldots, L$ and $\int_{\mathcal{V}}\psi_l(v) \psi_{l'}(v) d v = 0$ for $l \neq l'$. In practice, the orthonormal basis functions can be constructed from the eigenfunctions of a GP covariance kernel. One example is the modified squared-exponential covariance kernel, 
\begin{equation}
\label{kappa:se}
    \tilde{\kappa}(v_1,v_2) = \exp\{-a(\|v_1\|^2 + \|v_2\|^2) - b||v_1 - v_2||_2^2\},\quad \forall v_1,v_2\in \mathcal{V},\quad a > 0, \quad b>0.
\end{equation}
Here $||\cdot||_2$ denotes the Euclidean norm. For a GP with mean zero and covariance kernel $\tilde{\kappa}$,  the hyper-parameter $a$ in $\tilde{\kappa}$ controls the decay rate of its variance. The hyperparameter $b$ controls its smoothness; the smaller the value of $b$, the process becomes smoother. Note that the modified square kernel becomes a standard exponential kernel when $a = 0$. In our study, we set the hyperparameter $a$ with a small value (0.01) and estimate $b$ by averaging the estimated smoothing parameters of fitting GP models to the collected image outcomes of all ABCD subjects using the \texttt{GPfit} package in \texttt{R} \citep{macdonald2015gpfit}. Details of the hyperparameter estimation procedure are in the supplementary materials \citep{supplementary}. The choice of kernel form and distance measures should depend on the imaging data structure. For example, the appropriate distance measure for diffusion MRI data should be fiber length instead of Euclidean distance. And if we want to model a less smooth imaging data structure, we may consider using a more flexible kernel form such as the Gamma-exponential kernel \citep{williams2006gaussian}.

For $k = 1,\ldots, K$, $j =0,\ldots p$, $s = 1,\ldots, S$ and $r = 1,\ldots, q$, we assume,
\begin{align}
\label{eq:svcf basis expansion}
&\alpha_{kj}(v) = \sum_{l=1}^L \theta^\alpha_{kjl} \psi_l(v),\quad \gamma_{s}(v) = \sum_{l=1}^L \theta^\gamma_{sl}\psi_l(v)\quad  \mbox{and}\quad 
\eta_r(v) = \sum_{l=1}^L \theta^\eta_{rl}\psi_l(v),\quad  \forall v\in\mathcal{V},
\end{align}
where $\theta^\alpha_{kjl}$, $\theta_{sl}^\gamma$, and $\theta^\eta_{rl}$ are unknown basis coefficients to be estimated.  
 With the same set of orthonormal basis functions $\{\psi_l(v)\}_{l=1}^L$, we construct the covariance kernel $\kappa$ in \eqref{mod:1st} as

\begin{align}\label{eq:kernel}
&\kappa(v_1,v_2) = \sum_{l=1}^L \lambda_l\psi_l(v_1)\psi_l(v_2),\quad \forall v_1,v_2 \in \mathcal{V},
\end{align}
where $\{\lambda_l\}_{l=1}^L$ are unknown parameters with $\lambda_l>0$, which ensures that the covariance kernel $\kappa$ is positive definite~\citep{ghosal2017fundamentals_nonparam_bayes}. 

Model (\ref{mod:1st}) thus becomes:
\begin{equation}
    \label{mod:2nd}
          y_i(v) = \sum_{k=1}^K\delta_{ik}\sum_{l=1}^L \sum_{j=0}^p x_{ij}\theta^\alpha_{kjl} \psi_l(v) + \sum_{r=1}^q\sum_{l=1}^L z_{ir}\theta^\eta_{rl}\psi_l(v)+ \sum_{s = 1}^S\sum_{l=1}^L u_{is}\theta^\gamma_{sl}\psi_l(v) + \epsilon_i(v),
\end{equation}
where the number of unknown parameters is on the order of $L$ and substantially smaller than the number of voxels, $d$. The standard imaging data such as the volumetric fMRI data, even with a low-resolution image, may have at least $d = 2\times 10^4$ voxels.

In practical applications, the value of hyperparameters in the modified squared-exponential kernel defined in 
\eqref{kappa:se}, such as the decay rate parameter $a$ and smoothness parameter $b$, determine the required number of basis functions $L$. This method is based on the total variance explained by the eigenfunctions. We specify $h$ through $L$ based on the variance contribution rate $R$ which is defined as $R = \sum_{l=1}^L e_l/\sum_{l=1}^{\infty} e_l$, where $e_l$ the $l$th largest eigenvalues of the covariance kernel. We choose the smallest $L$ such that $R\geq R_0$. Some existing work suggests that $R_0$ is around $60\%$ leads to satisfactory performance in fitting fMRI data~\citep{wu2022bayesian}. 
To estimate $R$ in practice, we approximate $\sum_{l=1}^{\infty} e_l$ by a summation of truncated series $\sum_{l=1}^{L'} e_l$, where $L'$ is the reference number of basis functions and is typically chosen smaller than the sample size. In the analysis of ABCD fMRI data, we set the reference number of basis $L' = 1140$ (corresponding to $h=17$) and $R_0 = 0.6$ and obtain $L=680$ which corresponds to $h = 14$. This setting provides a balance between computational costs and model fitting performance. In our ABCD study, we have discovered that setting $h=14$, resulting in $L=680$, provides a conservative approximation that allows for reasonable model computation and fitting performance. Further details on constructing Hermite polynomials can be found in the Supplementary Materials \citep{supplementary}.

\subsection{Parameter Estimation}
\label{model:est}
We first discuss how to obtain orthonormal basis functions $\left\{\psi_l(v)\right\}^L_{l=1}$ for dimension reduction. Second, we describe the proposed SEM algorithm for efficient parameter estimation. Third, we present our model selection criteria for diagnostics. Finally, we develop the inferential process for the group-specific coefficients.

\subsubsection{Orthonormal Basis Functions and Model Representation}
\label{ss:basis}

We construct $\{\psi_l(v)\}_{l=1}^L$ on the set of voxels $\mathcal{V} = \{v_m\}_{m=1}^d$ from  $L$ eigenfunctions, denoted as $\{\tilde{\psi}_l(v)\}_{l=1}^L$ of a pre-specified covariance kernel function on $\mathbb{R}^3$. 
First, we evaluate $\{\tilde{\psi}_l(v)\}_{l=1}^L$ on $\mathcal{V}$ and obtain the matrix $\tilde{\Psi} = \{\tilde{\psi}_l(v_m)\}_{L\times d}$. Second we perform the singular value decomposition:  $\tilde{\Psi} = \Psi \mathbf{D} \mathbf{V}^\top$. Here, $\Psi$ is a $d\times L$ rotation matrix of the functional eigenvector and satisfies the orthonormal condition, i.e.,  $\Psi^\top \Psi = I_L$, where $I_L$ is an $L\times L$ identity matrix. Finally, we use the orthonormal functions to specify the matrix $\Psi=\{\psi_l(v_m)\}_{d\times L}$ and reduce Model \eqref{mod:2nd} to an $L$-dimensional multivariate linear regression model with diagonal variance structure. The derivation is shown below.

Denote by $\mathbf{y}_i$ a $d$-dimensional vector of image outcome of individual $i$ on the set of voxels $\mathcal{V}$. The matrix form of  equation~\eqref{mod:2nd} is specified as
\begin{equation}
\label{eq:proposition 1 proof (1)}
   \mathbf{y}_i = \sum_{k =1}^K \delta_{ik}\mathbf{x}_i\boldsymbol{\theta}^\alpha_k\Psi^\top + \mathbf{z}_i\boldsymbol{\theta}^\eta\Psi^\top + \mathbf{u}_i \boldsymbol{\theta}^\gamma\Psi^\top + \boldsymbol{\epsilon}_i,\quad
       \boldsymbol{\epsilon}_i \sim \mathcal{N}(\mathbf{0},\mathbf{K}),\quad \mathbf{K} = \Psi \Lambda \Psi^\top,
\end{equation}
where $\boldsymbol{\epsilon}_i = \{\epsilon_i(v_1),\ldots, \epsilon_i(v_d)\}^{\top}$, $\mathbf{K} = \{\kappa(v_m,v_{m'})\}_{d\times d}$ and  $\Lambda = \mathrm{diag}(\lambda_1,\ldots, \lambda_L)$. Since the constant matrix $\Psi$ is a $d\times L$ orthonormal matrix, i.e., $\Psi^\top\Psi = I_L$, Model \eqref{eq:proposition 1 proof (1)} is equivalent to the following representation:
\begin{equation}
\label{eq:model matrix}
    \Tilde{\mathbf{y}}_i = \sum_{k =1}^K \delta_{ik}\mathbf{x}_i\boldsymbol{\theta}^\alpha_k + \mathbf{z}_i\boldsymbol{\theta}^\eta + \mathbf{u}_i \boldsymbol{\theta}^\gamma + \Tilde{\boldsymbol{\epsilon}}_i,
\end{equation}
where $\Tilde{\mathbf{y}}_i = \mathbf{y}_i\Psi$ is a $L$-dimensional vector, with the transformed random error term $\Tilde{\boldsymbol{\epsilon}}_i = \boldsymbol{\epsilon}_i\Psi \sim \mathcal{N}(\mathbf{0},\mathbf{K}')$ with the covariance matrix $\mathbf{K}' = \Psi^\top\Psi \Lambda \Psi^\top\Psi = \Lambda$. Therefore, Model \eqref{eq:model matrix} is a formulation of $L$-dimensional multivariate regression problem, with error terms $\Tilde{\boldsymbol{\epsilon}}_i$ follows a normal distribution with mean zero and variance $\Lambda$.

\subsubsection{Stochastic Expectation Maximization Algorithm}

We develop an SEM algorithm \citep{diebolt1996stochastic}
to estimate parameters in LASIR. The SEM algorithm includes an expectation step (\textit{E}-step), a maximization step (\textit{M}-step), and an additional stochastic step (\textit{S}-step) at each iteration. Unlike the EM algorithm, the stochastic step in SEM improves convergence to the global optimum, rather than the local optimum, when the target distribution is multimodal \citep{grun2008flexmix}.

Let $\Omega = \{\Lambda, \boldsymbol{\theta}^\gamma, \boldsymbol{\theta}^\eta\}$ be the collection of parameters that are constant across latent subgroups, in contrast to group-specific parameters $(\boldsymbol{\theta}^\alpha, \mathbf{w})$. To estimate $(\Omega, \boldsymbol{\theta}^\alpha, \mathbf{w})$ based on the data $\{(\Tilde{\mathbf{y}}_i, \mathbf{x}_i, \mathbf{z}_i, \mathbf{u}_i)\}_{i=1}^n$, we describe the SEM steps below, given the current parameter estimated in $t$-th iteration.

At the \textit{E}-step, the a-posteriori probability of individual $i$ being assigned to subgroup $k$, for $k = 1,\ldots, K$ and $i=1,\ldots, n$, is estimated by
  \[\hat{p}_{ik} = \dfrac{\mathrm{Pr}(\delta_{ik} = 1| \mathbf{w}^{(t)}, \mathbf{z}_i)f_L\left(\Tilde{\mathbf{y}}_i\Big|\mathbf{x}_i, \mathbf{z}_i, \mathbf{u}_i, \Omega^{(t)},\boldsymbol{\theta}_k^{\alpha(t)}\right)}{\sum_{c=1}^K\mathrm{Pr}(\delta_{ic} = 1| \mathbf{w}^{(t)}, \mathbf{z}_i)f_L\left(\Tilde{\mathbf{y}}_i\Big|\mathbf{x}_i, \mathbf{z}_i, \mathbf{u}_i,\Omega^{(t)}, \boldsymbol{\theta}_c^{\alpha(t)}\right)}.\]

At the \textit{S}-step, given $(\hat{p}_{ik})_{k=1}^K$ in the \textit{E}-step, SEM draws the latent subgroup indicator $\hat{\boldsymbol{\delta}}_i^{(t)}$ from a categorical distribution with values of 1/0:  
$\hat{\boldsymbol{\delta}}_i^{(t)} \sim \mathrm{Categorical}\left((\hat{p}_{ik})_{k=1}^K\right).$

At the \textit{M}-step, the parameters $(\Omega^{(t+1)}, \boldsymbol{\theta}^{\alpha(t+1)}, \mathbf{w}^{(t+1)})$ are estimated by maximizing the conditional target log-likelihood function $Q$, where 
    $$Q = \sum_{i=1}^n\sum_{k=1}^K\hat{\delta}_{ik}^{(t)}\left( \log f_L(\Tilde{\mathbf{y}}_i\Big|\mathbf{x}_i,\mathbf{u}_i,\mathbf{z}_i, \Omega^{(t+1)},\boldsymbol{\theta}_k^{\alpha(t+1)}) + \log\mathrm{Pr}(\delta_{ik} = 1| \mathbf{w}^{(t+1)}, \mathbf{z}_i)\right).$$
Specifically, $(\Omega^{(t+1)}, \boldsymbol{\theta}^{\alpha(t+1)})$ can be estimated via maximizing likelihood estimation (MLE) in a linear regression model, where we first regressed transformed image outcomes $\tilde{\mathbf{y}}_i$ on $\mathbf{u}_i$ and  $\mathbf{z}_i$ with fixed effects, and second we used the resulting residuals as the outcome and estimated group-specific basis coefficients $\boldsymbol{\theta}^\alpha_k$ of the exposure variable $\mathbf{x}_i$, separately for subgroup $k=1,\ldots, K$. To estimate $\hat{\mathbf{w}}^{(t+1)}$, we use MLE in a multinomial-logit regression model with $\hat{\boldsymbol{\delta}_i}^{(t)}$ as the discrete outcome and $\mathbf{z}_i$ as covariates.

The above SEM algorithm converges when the fluctuation of the observed $Q$ values falls below a pre-specified tolerance level. 

\subsubsection{Model Selection}
\label{subsec: model selection}
To select the number of subgroups $K$, we use the Bayesian information criterion ~\citep[BIC]{schwarz1978estimating}: 
$\mathrm{BIC}(K) = M\log(nL) - 2Q,$
where $M$ is the total number of unknown parameters. For each subgroup $k$, there are $L(p+1)$ group-specific coefficients in $\boldsymbol{\theta}^\alpha_k$ and $L$ diagonal variance parameters. The numbers of fixed parameters in $\boldsymbol{\theta}^\gamma$ and $\boldsymbol{\theta}^\eta$ are $SL$ and $qL$, respectively. The distribution of the multinomial-logit random weight $\mathbf{w}$ involves $(K-1)(q+1)$ parameters. Furthermore, there are $L$ variance parameters in the diagonal variance matrix $\Lambda$. Thus, the total number of unknown parameters in the model is $M= KL(p+1) + KL + (S+q)L + (K-1)(q+1) + L$.

\subsection{Inference on the Group-specific Coefficients}
\label{ss:model inference}
We are interested in making inferences on SVCs for the group-specific coefficients, i.e., $\left\{\alpha_{k}(\mathcal{V})\right\}_{k=1}^K$, where $\alpha_{k}(\mathcal{V}) = \{\alpha_{kj}(\mathcal{V})\}_{j=1}^p$ and can be represented by a linear combination of basis functions: $\alpha_{kj}(\mathcal{V})= \sum_{l=1}^L \theta^\alpha_{kjl} \psi_l(\mathcal{V})$, to examine heterogeneity in the associations between brain activities and individual characteristics.

In Model~\eqref{mod:2nd}, the group-specific SVC $\left\{\alpha_{k}(\mathcal{V})\right\}_{k=1}^K$ is conditionally independent of the fixed effect SVCs, $\gamma_s(\mathcal{V})$ and $\eta_r(\mathcal{V})$, for $s = 1,\ldots, S$ and $r = 1,\ldots, q$. To obtain the group-specific covariance matrix of $\alpha_{kj}(\mathcal{V})$, we can first obtain the information matrix of  $\boldsymbol{\theta}^\alpha_{kj} = \{\theta^\alpha_{kjl}\}_{l=1}^L$ for $k=1,\ldots,K$. Considering the basis coefficients $\boldsymbol{\theta}^\alpha_k$ are independent across subgroups, let $\mathcal{I}_k = \{n:\delta_{ik} = 1\}$ be the collection of individuals in the $k$-th subgroup,  and the distribution of estimated basis coefficients $\mathrm{vec}(\boldsymbol{\theta}^\alpha_{k})$ becomes:
$\mathrm{vec}(\boldsymbol{\theta}^\alpha_{k}) \sim \mathcal{N}\left(\mathbf{0},\Lambda\otimes (\mathbf{x}_{\mathcal{I}_k}^{\top} \mathbf{x}_{\mathcal{I}_{k}})^{-1} \right)$, where $\mathrm{vec}(\cdot)$ is vectorization operator.
With the empirical estimate of error covariance matrix $\hat{\Lambda}$, the covariance matrix of $\boldsymbol{\theta}^{\alpha}_k$ is $\mathrm{Cov}(\boldsymbol{\theta}^\alpha_k) = \hat{\Lambda}\otimes (\mathbf{x}_{\mathcal{I}_k}^{\top} \mathbf{x}_{\mathcal{I}_k})^{-1}$. Since $\hat{\alpha}_k(\mathcal{V})$ is a linear transformation of $\boldsymbol{\theta}^\alpha_k$, the variance of group-specific SVC parameters $\hat{\boldsymbol{\alpha}}_{k}(\mathcal{V})$ can be obtained by:
\begin{equation}
      \mathrm{Cov}\{\hat{\boldsymbol{\alpha}}_{k}(\mathcal{V})\} =\Psi \otimes \left\{\hat{\Lambda}\otimes (\mathbf{x}_{\mathcal{I}_k}^{\top} \mathbf{x}_{\mathcal{I}_k})^{-1}\right\}\otimes\Psi^\top.
\end{equation}

Here the diagonal elements of $\mathrm{Cov}\{\hat{\boldsymbol{\alpha}}_{k}(\mathcal{V})\}_{d(p+1) \times d(p+1)}$  are the variance estimates of $\hat{\alpha}_{kj}(v)$ at voxel $v$. With the variance estimates, we can test the null hypothesis $H_0: {\alpha}_{kj}(v) = 0$, for each $v$, with the Wald test statistics:
\begin{align}
    W_{kj}(v) &= \dfrac{|\hat{\alpha}_{kj}(v)|}{\sqrt{\mathrm{Var}\left(\hat{\alpha}_{kj}(v)\right)}}\sim \mathcal{N}(0, 1), \quad k=1,\ldots,K,\quad j = 0,\ldots, p.
\end{align}
With the Wald test statistics computed at each voxel $v$, we can obtain the mapping image of p-values. To control potential FDR, we perform multiple comparison corrections on the p-value image using the random field theory ~\citep{nichols2003fwe}. For interpretation purposes, we summarize the significant voxels at the functional region level based on certain parcellation.  

The computational codes are available for replication and public use in an R package on~\href{https://github.com/zikaiLin/lasir}{Github}.

\section{Simulations}
\label{s:simulation}
We conducted simulation studies to evaluate LASIR's performances on the subgroup identification and parameter estimation accuracy in comparison with existing subgroup identification algorithms. We considered two simulation designs of the imaging space, labeled as `cube shape` and `brain shape.`

\begin{figure}
    \includegraphics[width=5in]{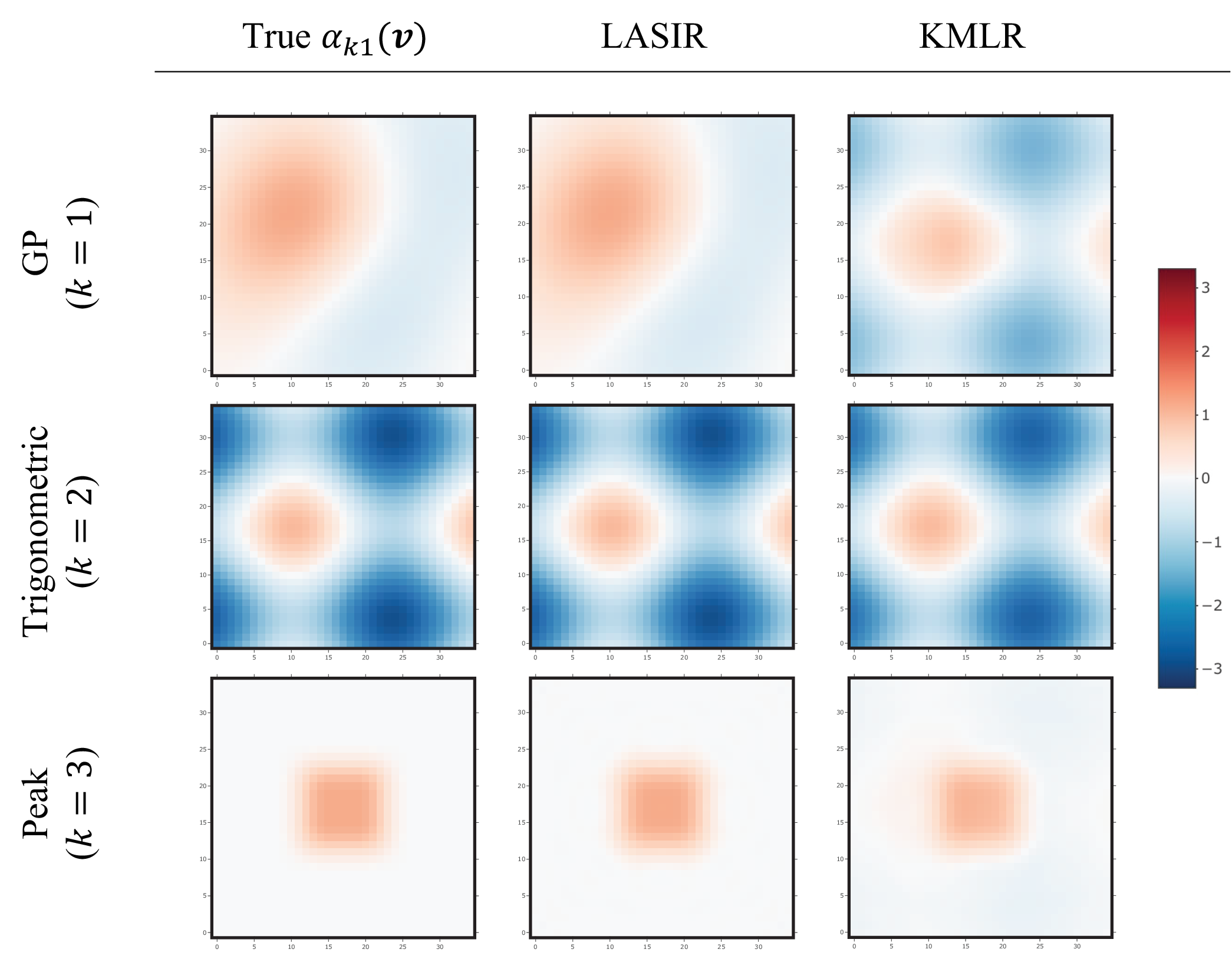}
\caption{ Cube shape: comparison of simulated and estimated values of group-specific spatially varying coefficient (SVCs) $\{\alpha_{k1}(\mathcal{V})\}_{k=1}^3$. For subgroup $k=1$ (top row), $\alpha_{11}(\mathcal{V})$ is generated from a Gaussian process with mean zero and squared-exponential kernel. For subgroup $k=2$ (middle row), $\alpha_{21}(\mathcal{V})$ is generated from a trigonometric function. For $k=3$ (bottom row), $\alpha_{31}(\mathcal{V})$ is sparse and has non-zero values only within the center area. The three columns display the true values, LAtent Subgroup Image-on Scalar (LASIR) estimates, and K-Means with Linear Regression (KMLR) estimates, respectively. The 13th slice on the 3rd dimension is shown (with sample size $n = 2000$ and dimension $d = 35^3$). }
     \label{f:simu_results}
\end{figure}

First, we discuss the generation process of a simulation study in the 3D cube. For each individual $i$, we generated a 3D image ($v\in \mathbb{R}^3$) with different dimensions: $d=25^3$ and $d = 35^3$. The scalar exposure variable $x_i$ and control variable $z_i$ were generated from $x_i\sim \mathcal{N}(0, 1)$ and $z_i \sim \mathcal{N}(0, 2)$, respectively. The intercepts $\alpha_{0}(\mathcal{V})$ were simulated from $\mathcal{GP}(\mathbf{0}, \kappa_0)$, where $\kappa_0$ was the modified squared-exponential kernel covariance matrix defined in \eqref{kappa:se}, with a smoothness parameter $b = 2$ and decay rate parameter $a = 0.01$.  We introduced three subgroups with different spatial structures (i.e., $K=3$) and group-specific coefficients $\alpha_{k1}(\mathcal{V})$'s: 1) continuous spatial structure with a normal distribution $\alpha_{11}(v) \sim\mathcal{N}_d(0, \kappa_0), 2)$ continuous spatial structure with a trigonometric function $\alpha_{21}(v) =  \sin(4v_x) + \cos(4v_y) - \sin(4v_z)$, where $(v_x, v_y, v_z)$ are coordinates of a 3D voxel, and 3) sparse discrete structure that only had signal within the center area within the smoothed boundary for $\alpha_{31}(v)$. The visualization of true main effect SVCs is presented in the left column of Figure \ref{f:simu_results}, labeled with cube shape.

The latent subgroup indicator $\delta_{ik}$ for individual $i$ was sampled from a categorical distribution with probability $\mathrm{Pr}(\delta_{ik} = 1|\mathbf{w}_k, z_i)$, for $k=1,\ldots,3$,
where $\mathbf{w}_1 = [-0.6,1]^\top$, $\mathbf{w}_2 = [0.5,1]^\top$, and $\mathbf{w}_3 = [0,0]^\top$.
We simulated the site-specific coefficients $\gamma_{s}(v) \sim \mathcal{N}(0, 0.2^2)$ ($s =1,\ldots, S$), and the coefficient of the control variable $z_i$: $ \eta_1(\mathcal{V}) \sim \mathcal{N}_d(0, 0.2^2\mathbf{I}) $. We specified two different values as the standard deviation of the random noise $\epsilon_i(v)$ at $\sigma(v)\in \{1,4\}$ to introduce different signal-to-noise levels of the imaging outcome. Finally, the outcome $y_i(v)$ of individual $i$ at voxel $v$ was generated by $y_i(v) \sim \mathcal{N}_d\left(\mu_i(v), \sigma^2(v)\right)$, where 
   $ \mu_i(v) = \beta_{i0}(v) + \beta_{i1}(v)x_i + \sum_{s=1}^S\gamma_{s}(v)u_{is} + z_i\eta_1(v) $ and $\beta_{i1}(v) = \sum_{k=1}^K\delta_{ik}\alpha_{k1}(v)$. Note that the simulated errors were independent; nevertheless, we applied the basis expansion approach as the general computational strategy for the LASIR.

\begin{figure}
    \centering
    \includegraphics[width=\textwidth]{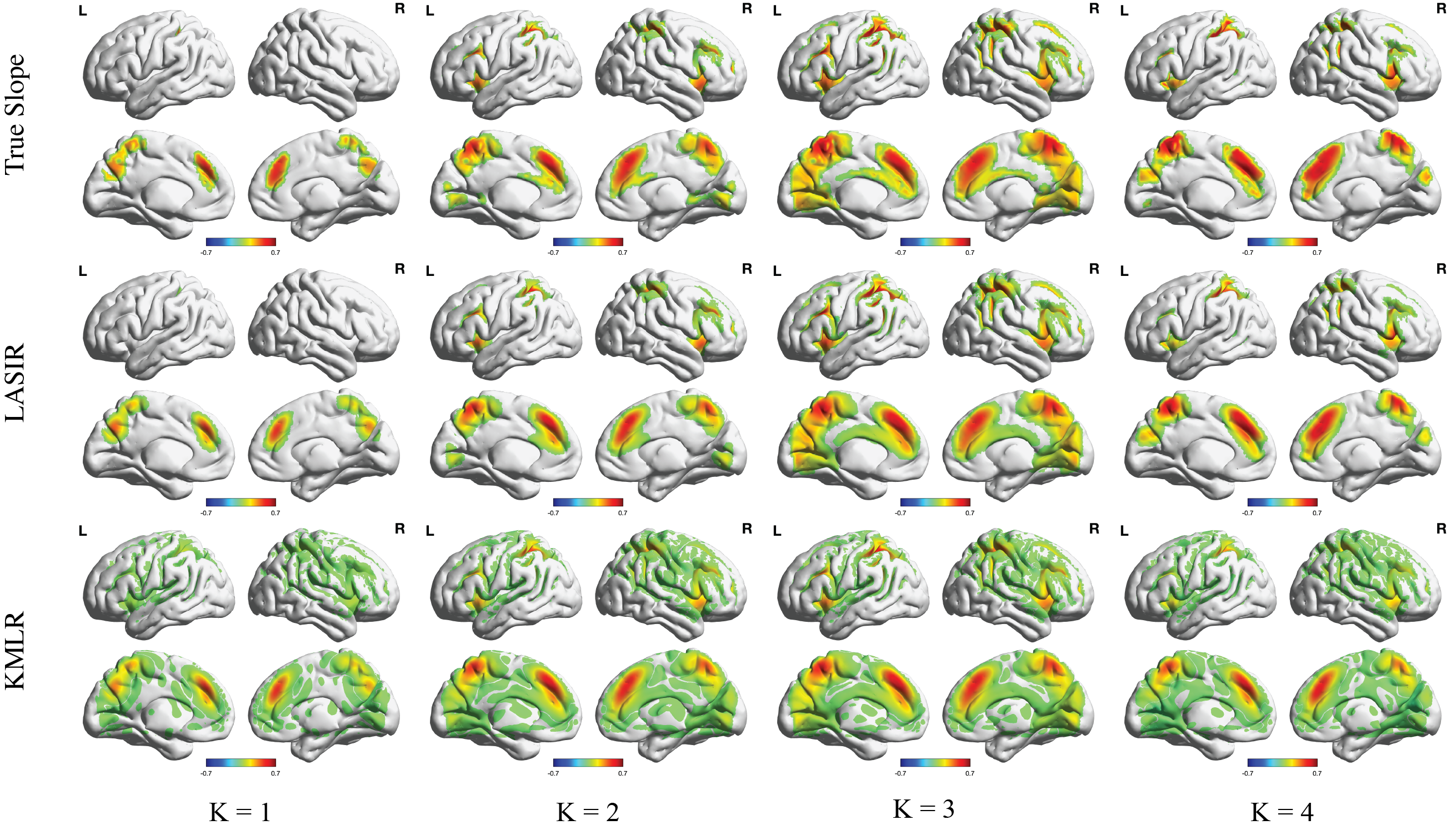}
      \caption{Brain shape: comparison of simulated and estimated values of group-specific spatially varying coefficient (SVCs) as group-specific $\{\alpha_{k1}(\mathcal{V})\}_{k=1}^4$. The three panels display the true values, LAtent Subgroup Image-on Scalar (LASIR) estimates, and K-Means with Linear Regression (KMLR) estimates, respectively (sample size $n = 2000$). }
    \label{fig:simulation-brain}
\end{figure}

We also perform a simulation study that mimics our real data analysis, labeled with `brain shape`. We used the standard 3mm fMRI volumetric data in MNI space as our collection of voxels $\mathcal{V}$. The scalar exposure variable $x_i$ was generated from $x_i\sim \mathcal{N}(0,1)$. We specified two binary control variables  $z_{i1}\sim \mathrm{Bernoulli}(0.4)$ and $z_{i2}\sim \mathrm{Bernoulli}(0.75)$, respectively. To mimic our real data analysis, we specified $K = 4$ different subgroups with group-specific slopes similar to the estimates obtained from our real data analysis. We adjusted the $\alpha_{k1}(\mathcal{V})$  to achieve a signal-to-noise ratio of 0.9. To maintain the sparse structure, we set $\alpha_{k1}(v) = 0$ for voxel $v$ where $|\alpha_{k1}(v)| < 0.03$ for all $k = 1,\ldots, 4$. The group-specific intercepts were simulated from $\mathcal{GP}(\mathbf{0}, \kappa_0)$, where $\kappa_0$ was the modified squared-exponential kernel covariance matrix with a smoothness parameter $b = 2$ and decay rate parameter $a = 0.01$. The brain mapping visualization of $\{\alpha_{k1}(\mathcal{V})\}_{k=1}^4$ is shown in Figure \ref{fig:simulation-brain}. 
To further illustrate the practicality of our LASIR model on real data, we changed the standard deviation of the error term $\sigma(v)\in \{0.05, 0.01\}$. We set up the signal-to-noise level of our simulation studies based on our analysis of real data, in which we estimated the standard deviation of residuals $\sigma(v)$ to range from 0 to 0.2, with an average of 0.04.
   
We constructed $455$ orthonormal basis functions (highest degree of Hermite polynomials $h= 12$) for simulated images with dimension $d = 25^3$ and $680$ basis functions ($h=14$) for images with $d = 35^3$. For simulated images with a similar data structure to the real fMRI data, we constructed 968 basis functions (highest degree of Hermite polynomials $h = 16$).  For each of the two image dimensions, we simulated $n = 1000$ or $n= 2000$ individuals as one dataset. Under the same design, we repeatedly generated $50$ datasets under the cube shape simulation and 30 datasets for the brain-like simulation.

We compared our method with the K-Means algorithm in combination with Linear Regressions (KMLR, \citet{hartigan1979ak}). KMLR first applied K-Means classification to the outcome image $\left(\mathbf{y}_i\right)_{i=1}^n$ and obtained the subgroup indicator $\boldsymbol{\delta}_i$ for each individual. Second, the coefficient and covariance parameters were estimated in the same way as the M-step of LASIR. We iteratively updated all parameters until convergence.

We also compared the performance of LASIR and KMLR with the traditional SVCM approach without subgroup identification. To estimate the accuracy of reconstructed SVCs, we calculated the mean squared error (MSE) of individual-specific parameters $\beta_{ij}(\mathcal{V})$'s, instead of $\alpha_{kj}(\mathcal{V})$'s.  To evaluate the classification accuracy of LASIR and KMLR, we utilized the normalized mutual information (NMI, \citet{rousson2003active}) to quantify the mutual information between two cluster labels. Let $\mathcal{C}_1$ and $\mathcal{C}_2$ be two sets of disjoint cluster labels of all observations, where $\mathrm{NMI}(\mathcal{C}_1,\mathcal{C}_2) = 0$ indicates that the two clusters share no labels with each other, and $\mathrm{NMI}(\mathcal{C}_1,\mathcal{C}_2) = 1$ indicates that these two labels are exactly the same. We calculated the MSE of the SVCs and NMI between the true and estimated cluster labels for each replication and then averaged over all repetitions to obtain the overall MSE and NMI. Table \ref{tab:simu_res} shows that in the cube shape simulation study LASIR is able to recover the group-specific coefficients $\alpha_k(\mathcal{V})$'s and individual-specific coefficients $\beta_i(\mathcal{V})$'s with smaller MSEs compared to KMLR. LASIR also outperforms KMLR in clustering accuracy across all simulation scenarios. In the brain structure simulation study, shown in Table~\ref{tab:simu-res-brain}, our LASIR method outperforms the KMLR in both clustering accuracy and robustness of significant voxels detection. We have evaluated the Type I error rate and power obtained from different approaches, and LASIR's improvement in clustering accuracy and controlling for Type I error is apparent over KMLR.  

\begin{table}[h]
  \caption{Average normalized mutual information (NMI), mean squared error (MSE) of estimates, Type I error (\%) and power of detection generated by K-Means with Linear Regression (KMLR), LAtent Subgroup Image-on Scalar (LASIR) for group-specific coefficients $\{\alpha_k(\mathcal{V})\}_{k=1}^4$, shared coefficients $\{\gamma_s(\mathcal{V})\}_{s=1}^{21}$, and $\eta_1(\mathcal{V})$. The NMI and MSE are computed over 30 simulated datasets.}
\begin{tabular}{lrrrr}
\hline
&\multicolumn{2}{c}{$\sigma_i(v) = 0.05$} & \multicolumn{2}{c}{$\sigma_i(v) = 0.1$} \\
\hline
$n$ & $1000$ & $2000$  & $1000$ & $ 2000$ \\
\hline
\multicolumn{5}{c}{Clustering Accuracy (Average NMI)}\\[2mm] 
\hline
KMLR  & 0.239 & 0.264  & 0.235 & 0.266  \\ 
LASIR  & 0.995   &0.997  & 0.992 & 0.995     \\[2mm] 
\hline
\multicolumn{5}{c}{MSE of  $\{\alpha_k(\mathcal{V})\}_{k=1}^4$ (Unit: $\times 10^2$) }\\[2mm] 
\hline
KMLR &  0.604 & 0.626   & 0.626 &  0.621   \\
LASIR  & 0.654  & 0.514  & 0.460 & 0.450     \\[2mm]
\hline 
\multicolumn{5}{c}{MSE of constant coefficients $\{\gamma_s(\mathcal{V})\}_{s=1}^{21}$ }\\[2mm] 
\hline
KMLR& 0.001 & 0.001 &0.001& 0.001 \\ 
LASIR & 0.002 & 0.001 & 0.001 & 0.001 \\ 
\hline
\multicolumn{5}{c}{MSE of constant coefficients $\eta_1(\mathcal{V})$ }\\[2mm] 
\hline
KMLR& 0.040 & 0.040 & 0.039 & 0.039  \\ 
LASIR & 0.039 & 0.039 & 0.039 & 0.039 \\ 
\hline
\multicolumn{5}{c}{Power (\%) }\\[2mm] 
\hline
KMLR& 89.6 & 94.9 & 89.3 & 95.2  \\ 
LASIR & 59.3 & 82.4 & 94.4 & 97.8 \\ 
\hline
\multicolumn{5}{c}{Type I Error (\%) }\\[2mm] 
\hline
KMLR  & 26.2& 37.4 & 27.0 & 36.9  \\ 
LASIR & 0.8 & 2.0 & 4.5 & 7.5 \\ 
\hline
\label{tab:simu-res-brain}
\end{tabular}
\end{table}

\begin{table}[h]
  \caption{Average normalized mutual information (NMI) and mean squared error (MSE) of estimates generated by K-Means with Linear Regression (KMLR), LAtent Subgroup Image-on Scalar (LASIR), and Spatial Varying Coefficient Model (SVCM) for group-specific coefficients $\{\alpha_k(\mathcal{V})\}_{k=1}^3$, shared coefficients $\{\gamma_s(\mathcal{V})\}_{s=1}^{21}$, $\eta_1(\mathcal{V})$,  and individual-specific coefficients $\{\beta_i(\mathcal{V})\}_{i=1}^n$. The NMI and MSE are computed over 50 simulated datasets.}
\begin{tabular}{lrrrrrrrr}
\hline
&\multicolumn{4}{c}{$d = 25^3$} &\multicolumn{4}{c}{$d = 35^3$}\\
\hline
&\multicolumn{2}{c}{$\sigma_i(v) = 1$} & \multicolumn{2}{c}{$\sigma_i(v) = 4$} & \multicolumn{2}{c}{$\sigma_i(v) = 1$}  & \multicolumn{2}{c}{$\sigma_i(v) = 4$}\\
\hline
$n$ & $1000$ & $2000$ & $1000$ & $2000$  & $1000$ & $2000$  & $1000$ & $ 2000$ \\
\hline
\multicolumn{9}{c}{Clustering Accuracy (Average NMI)}\\[2mm] 
\hline
KMLR & 0.305 & 0.309 & 0.302 & 0.307 & 0.305 & 0.308 & 0.302 & 0.304 \\ 
LASIR   & 0.964 & 0.963 & 0.856 & 0.878 & 0.971 & 0.974 & 0.891 & 0.923 \\[2mm] 
\hline
\multicolumn{9}{c}{MSE of group-specific main effect coefficients $\{\alpha_k(\mathcal{V})\}_{k=1}^3$ (unit: $\times 10^3$)}\\[2mm] 
\hline
KMLR & 11.463  & 9.019   & 15.138 & 10.903   &  11.769 & 9.419  & 13.253   & 9.680 \\
LASIR  & 0.176  & 0.113  & 1.934 & 1.070    & 0.123  &  0.083    &  1.039 & 0.435 \\[2mm]
\hline
\multicolumn{9}{c}{MSE of individual-specific coefficients $\{\beta_i(\mathcal{V})\}_{i=1}^n$ (unit: $\times 10^3$)}\\[2mm] 
\hline
KMLR& 14.506 & 11.209 & 18.977 & 13.214 & 14.734 & 11.688 & 16.519 & 11.860\\ 
LASIR  & 0.144 & 0.097 & 1.534 & 0.817 & 0.106 & 0.076 & 0.840 & 0.373 \\ 
SVCM  & 18.281 & 13.207 & 30.190 & 20.419 & 17.619 & 10.068 & 25.039 & 12.832 \\[2mm] 
\hline
\multicolumn{9}{c}{MSE of site-specific coefficients $\{\gamma_s(\mathcal{V})\}_{s=1}^{21}$ }\\[2mm] 
\hline
KMLR& 0.145 & 0.128 & 0.163 & 0.137 & 0.144 & 0.128 & 0.154 & 0.131 \\ 
LASIR  & 0.041 & 0.040 & 0.059 & 0.049 &0.041 & 0.040 & 0.051 & 0.043 \\ 
SVCM  & 0.056 & 0.049 & 0.074 & 0.058 & 0.057 & 0.048 & 0.066 & 0.051 \\[2mm] 
\hline 
\multicolumn{9}{c}{MSE of constant coefficients $\eta_1(\mathcal{V})$ }\\[2mm] 
\hline
KMLR& 0.039 & 0.039 & 0.039 & 0.039 & 0.040 & 0.040 & 0.040 & 0.040 \\ 
LASIR & 0.039 & 0.039 & 0.039 & 0.039 & 0.039 & 0.039 & 0.040 & 0.040\\ 
SVCM  &  0.039 & 0.039 & 0.039 & 0.039 & 0.040 & 0.039 & 0.040 & 0.040\\
\hline
\label{tab:simu_res}
\end{tabular}
\end{table}

We also verified the accuracy of LASIR model selection via BIC in the case when there was only one group existing in the dataset. When the true value is $K= 1$, Model~\eqref{mod:2nd} reduces to SVCM. We simulated $n = 1000$ images with dimension $d = 25^3$, and the group-specific intercept and slopes were set as the same $\alpha_{20}(v)$ and $\alpha_{21}(v)$ as those in cube shape simulation study. The LASIR selected $K=1$ as the optimal number of subgroups for all 50 simulated datasets. 

\section{Application to the ABCD study}
\label{ss:rda}

The ABCD study aims to study the U.S. children's brain development and health~\citep{casey2018adolescent}. The mapping structure between the brain activity and cognitive ability measures is not well understood, which could be moderated by many risk factors~\citep{sternberg2002general}. Analyzing the ABCD baseline data, we applied the proposed LASIR to take into account population heterogeneity and identify subgroups in association studies with inferential validity and substantive meanings.

Our analysis focused on the 2-Back versus 0-Back contrast maps derived from the working memory task-based fMRI data from the ABCD Study Curated Annual Release 1.0~\citep{ewing2018implications}. Designed to engage memory and emotion regulation processes, the 2-Back fMRI data refer to tasks with high memory conditions, and 0-Back refers to low memory conditions \citep{hagler2019ABCDpreproc}. The contrast maps were generated by standardizing the difference between the 2-Back fMRI data and the 0-Back fMRI data. Images with unusable T1w images, poor registration/normalization quality, or too much head motion, were excluded during the image preprocessing stage, resulting in 2,021 individuals for our analysis \citep{sripada2020prediction}. We treated the 2-Back v.s. 0-Back contrast image as the outcome $y_i(\mathcal{V})$ and  the total composite score of cognitive measure ($g$-factor) as the exposure variable of interest $x_{i}$, (i.e., $p=1$). We included age, family size, parental marital status, race/ethnicity, gender, parental education levels, and household income as control variables $\mathbf{z}_i$ with fixed effects after dummy coding of factors ($q=13$). Our analysis also included the study site indicator for each individual to model the site-specific fixed effect ($S = 21$).
 
\subsection{Image Preprocessing}

The neuroimaging data were preprocessed by a standard procedure described in \citep{hagler2019ABCDpreproc}. The images were acquired through standard scan sessions (2.4mm isotropic, TR = 800ms). The fMRI acquisitions use multi-band EPI with slice acceleration factor 6 and fieldmap scans for the $\mathrm{B}_0$ distortion correction. Head motion was corrected by registering frames to the first using AFNI's \texttt{3dvolreg} \citep{cox1996afni}, $\mathrm{B}_0$ distortions were corrected using the reversing polarity method described in \citep{holland2010echo}. To correct for between-scan motion, the reference scan was chosen to be the one nearest to the middle of fMRI scans for each precipitant, and then each scan was resampled with cubic interpolation into alignment with this reference scan. Registration was also done between spin-echo, $\mathrm{B}_0$ calibration scans, and $\mathrm{T}_1$w images using mutual information \citep{wells1996multi}. We performed additional image registrations and alignment via \texttt{fslr} package in \texttt{R}~\citep{muschelli2015fslr}. 

\subsection{Model Specification}
For the LASIR setup, we assigned the modified squared-exponential kernel~\eqref{kappa:se} with hyperparameters $b = 200$ and $a = 0.01$. With the reference number of basis $L'= 1140$ and the minimum variance contribution rate $R_0 = 60$\%, we have a total of $L = 680$ orthogonal basis functions in the orthogonal space. This specification corresponds to the degree of Hermite polynomial $h =  14$~\citep{BayesGPfit}. The outcome image of each individual $\mathbf{y}_i$ ($i=1,\ldots,2,021$) was mapped onto the orthogonal space. The number of subgroups was selected by minimizing BIC from eight candidate numbers ranging from one to eight. To reduce the impact of the initial values on the SEM algorithm, we repeatedly ran LASIR 20 times for each candidate model with different seeds and random initial values generated from a standard normal distribution. The SEM algorithm finished in 12 minutes and 56 seconds for in total of 20 runs under iMac Pro 2017, 3.2 GHz 8-core Intel Xeon CPU. The selected model converged after 11 iterations, and the scatter plot of observed log-likelihood over iterations was included in the Supplementary Materials \citep{supplementary}. We made statistical inferences based on covariance estimates and performed multiple comparison corrections to control the FDR. To facilitate the interpretation of the results, we have employed the Power 264 parcellation~\citep{power2011functional} to partition the brain volume into functional regions, and present the regions that contain statistically significant voxel-level findings. We also fitted SVCM for comparison.

\subsection{Results}
\label{ss:RDA Results}

\begin{figure}[htb]
     \centering
        \includegraphics[width=0.95\textwidth]{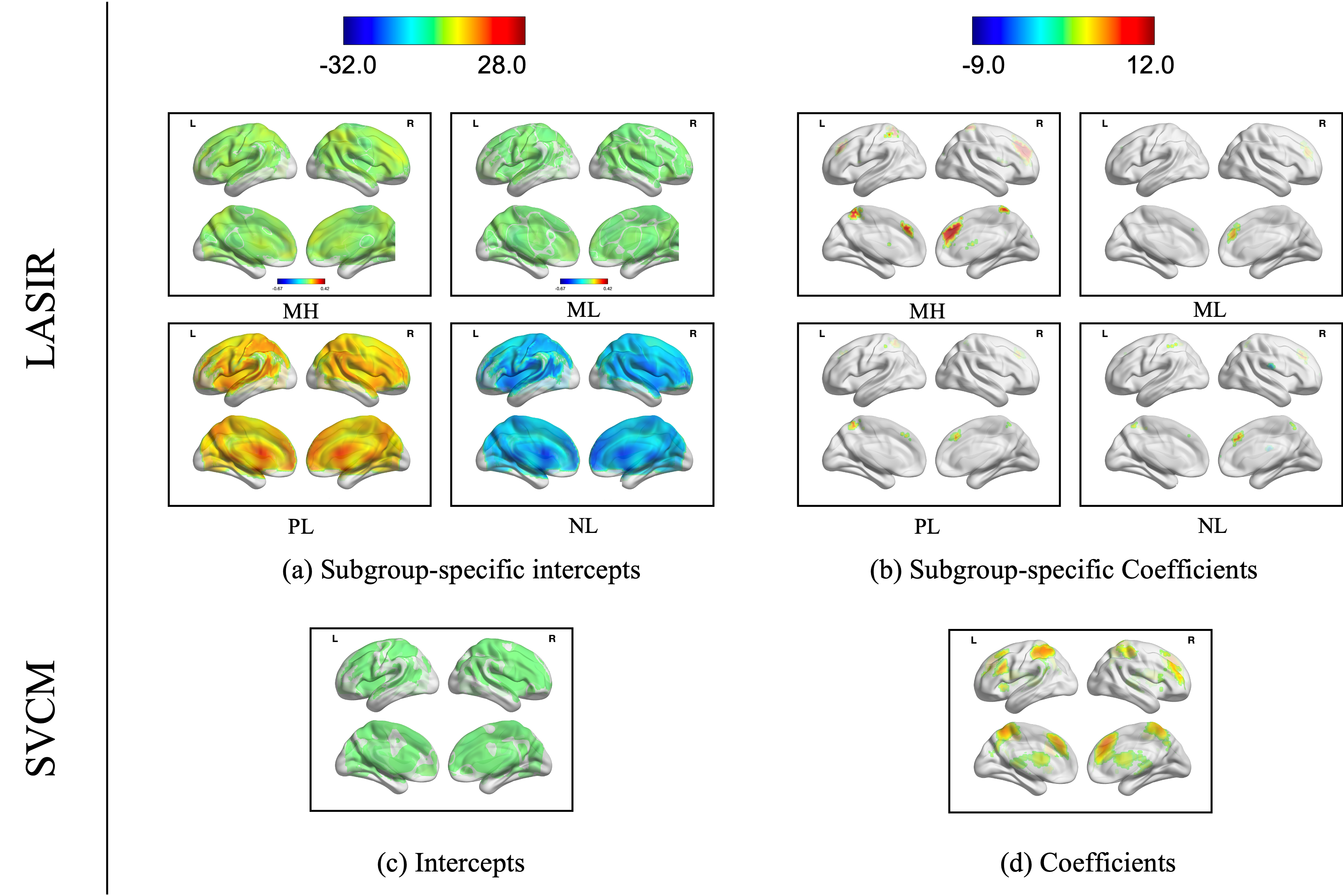}
               \caption{Effect size estimates of associations between imaging data under 2-Back v.s. 0-Back contrast and the g-factor, based on LAtent Subgroup Image-on Scalar (LASIR) and Spatial Varying Coefficient Model (SVCM) approaches. For estimation given by LASIR, group-specific intercepts $\left\{\alpha_{k0}(v)\right\}_{k=1}^4$ and associations with the $g$-factor $\left\{\alpha_{k1}(v)\right\}_{k=1}^4$ are displayed. All significant voxels displayed were corrected by FDR at a significance level $0.05$. }
     \label{f:g_res_brain}
\end{figure}

The LASIR identifies four subgroups among $n=2,021$ baseline participants in the ABCD study. According to estimated brain activity patterns and main effect SVCs, we refer to the four subgroups as: Medium contrast High association (MH) with estimated size $n_{\mathrm{MH}} = 243$, Medium contrast Low association (ML) with $n_{\mathrm{ML}} = 769  $, Positive contrast Low association (PL) with $n_{\mathrm{PL}}=  552$, and Negative contrast Low association (NL) with $n_{\mathrm{NL}}= 457$. Here the ``contrast" refers to the mean difference in brain activity patterns between the 2-Back and the 0-Back working memory tasks. The "association" refers to the association between the $g$-factor and the working memory brain activity. Figure \ref{f:g_res_brain} displays the brain activities for 2-Back minus 0-Back contrast and the significant associations between brain activities during 2-Back minus 0-Back contrast and the $g$-factor. The p-values were FDR-corrected with a probability threshold of $0.05$. The brain mapping of site-specific SVCs ($\{\gamma_s(\mathcal{V})\}_{s=1}^{21}$) and fixed control effect SVCs ($\{\eta_r(\mathcal{V})\}_{r=1}^{13}$) are presented in the Supplementary Materials \citep{supplementary}.

As illustrated in Figure \ref{f:g_res_brain}, in subgroups MH  and PL,  brain activities are positively related to $g$-factor in frontal-parietal task control networks (FPTC; superior frontal medial gyrus), default mode network (DMN; superior frontal medial gyrus; anterior cingulate gyri), salience network (SAL; anterior cingulate gyri, medial cingulate gyri), and sensory/somatomotor networks (SMN; precuneus; postcentral gyrus). For individuals in subgroup ML, only brain activities in DMN  have significant associations with the $g$-factor. For individuals in the NL subgroup, most brain activities are unrelated to the g-factor. Significant associations are only detected for a small number of voxels, among which the ones in the SMN show negative associations in contrast to those in the DMN with positive associations. Such findings are in line with previous studies on disparities in children's cognitive functions, which also identify group differences in associations between cognitive performance and brain activities in cognitive/working memory-related functional networks \citep{knoll2012leftprefrontalcortex, sherman2014development}. 
However, instead of treating subgroups as latent classes and taking the spatial dependency of brain imaging data into account, these studies first define subgroups based on hypotheses and then carry out voxel-wise one-sample t-tests. 
For subgroups PL and NL, the group differences are primarily captured by brain activities for the 2-Back minus 0-Back contrast; positive contrast (2-Back $>$ 0-Back) is observed in subgroup PL, while reversed contrast (2-Back $<$ 0-Back) is observed for individuals in subgroup NL. This finding provides evidence of potential differences in activation patterns of the working memory network between these two subgroups \citep{egli2018identification}.
 
Table~\ref{t:demographic} summarizes the sociodemographic decomposition across identified subgroups. For each sociodemographic factor, the significance of associations between pairwise subgroups’ contingency tables is assessed by Fisher’s exact tests, where the null hypothesis is that the two subgroups have the same distribution of the given sociodemographic variable. We have ignored the estimated variance in the latent group during the group comparison, which is a limitation of this analysis. In general, different subgroups present different decompositions of race, financial stress, and parental education. Our results are consistent with previous findings in the ABCD study \citep{assari2021parental}, which suggests that racial profiles  (specifically Black and White) and childhood poverty are associated with some altered function in brain regions related to executive and reward networks. These sociodemographic risk factors and regions of interest have also been considered to be significant predictors of subgroup allocation in our results. 

Specifically, subgroup MH has a larger portion of Black individuals (11.9\%) and Hispanic individuals (28.0\%) than all other subgroups. And compared to other subgroups, adolescents in subgroup MH are more likely to be from less-educated families, with only  53.9\% of their parents having Bachelor's degrees or above. Racial discrimination experiences have been shown to affect a variety of brain regions, such as the anterior cingulate and frontal medial \citep{clark2018experiences}. Here the disparities in racial profile and the income level of subgroup MH detected by LASIR are consistent with previous findings \citep{clark2018experiences}, where social discrimination and marginalization associated with race and income are also identified as risk factors for altering the brain functions in executive networks. However, our results  provide convincing statistical evidence of heterogeneity in brain activity and $g$-factor association.

Furthermore, adolescents in subgroup PL tend to be from highly educated and less financially stressful families, with 67\% of their parents having at least Bachelor's degrees as the highest parental education level and 47.8\% having household income higher than \$100K. Compared to subgroup NL, adolescents in subgroup ML are more likely to be the only child in their families (77.2\%). Our results do not indicate any significant distributional differences from other sociodemographic variables.

\begin{table}[h!]
 \caption{Descriptive summary of sociodemographic across identified subgroups, labeled as MH, ML, PL and NL. }
 \small
 \def\~{\hphantom{0}}
  \resizebox{0.95\textwidth}{!}{  
\begin{tabular}{llllll}
 \toprule
 \textbf{Category} & \textbf{All} & \textbf{MH} ($n_{\mathrm{MH}} = 243 $) &  \textbf{ML}  ($n_{\mathrm{ML}} =  769 $) & \textbf{PL} ($n_{\mathrm{PL}} = 552 $) & \textbf{NL} ($n_{\mathrm{NL}} =  457$)  \\ 
  \midrule
  \multicolumn{6}{l}{\textbf{Race/Ethnicity}}\\
 \hline
 Other+Asian& 210 (10.4\%) & 24 (9.9\%) & 100 (13.0\%) & 48 (8.7\%) & 38 (8.3\%) \\ 
  Black  & 147 (7.3\%)& 29 (11.9\%) & 55 (7.2\%) & 36 (6.5\%) & 27 (5.9\%) \\ 
  White  & 1237 (61.2\%) & 122 (50.2\%) & 462 (60.1\%) & 351 (63.6\%) & 302 (66.1\%) \\ 
  Hispanic & 427 (21.1\%)& 68 (28.0\%) & 152 (19.8\%) & 117 (21.2\%) & 90 (19.7\%) \\
 \midrule
    \multicolumn{6}{l}{\textbf{Parental Education}}\\
    \hline
    Post Graduate& 749 (37.1\%)& 68 (28.0\%) & 303 (39.4\%) & 200 (36.2\%) & 178 (38.9\%) \\ 
  Bachelors & 576 (28.5\%)& 63 (25.9\%) & 212 (27.6\%) & 187 (33.9\%) & 114 (24.9\%) \\
  College & 504 (24.9\%)& 74 (30.5\%) & 191 (24.8\%) & 109 (19.7\%) & 130 (28.4\%) \\
  High School/GED & 125 (6.2\%)& 30 (12.3\%) & 40 (5.2\%) & 28 (5.1\%) & 27 (5.9\%) \\ 
  High School & 67 (3.3\%)& 8 (3.3\%) & 23 (3.0\%) & 28 (5.1\%) & 8 (1.8\%) \\ 
    \midrule
   \multicolumn{6}{l}{\textbf{Household Income}} \\ 
   \hline 
   100K & 850 (42.1\%) & 90 (37.0\%) & 314 (40.8\%) & 264 (47.8\%) & 182 (39.8\%) \\ 
  50K-100K &600 (29.7\%)& 66 (27.2\%) & 233 (30.3\%) & 144 (26.1\%) & 157 (34.4\%) \\ 
  $<$50K & 571 (28.3\%)&87 (35.8\%) & 222 (28.9\%) & 144 (26.1\%) & 118 (25.8\%) \\ 
   \midrule
   \multicolumn{6}{l}{\textbf{Parental Marital Status}} \\ 
   \hline 
   Married & 1482 (73.3\%)& 166 (68.3\%) & 582 (75.7\%) & 405 (73.4\%) & 329 (72.0\%) \\
  Single & 539 (26.7\%)& 77 (31.7\%) & 187 (24.3\%) & 147 (26.6\%) & 128 (28.0\%) \\ 
 \midrule
   \multicolumn{6}{l}{\textbf{Only Child}}\\ 
   \hline
  Yes & 1504 (74.4\%)& 186 (76.5\%) & 594 (77.2\%) & 403 (73.0\%) & 321 (70.2\%) \\ 
  No & 517 (25.6\%)& 429 (23.5\%) & 175 (22.8\%) & 149 (27.0\%) & 136 (29.8\%) \\ 
    \hline
    \multicolumn{6}{l}{\textbf{Age}} \\ 
    \hline
    $\geq$10 & 1178 (58.3\%)& 141 (58.0\%) & 443 (57.6\%) & 318 (57.6\%) & 276 (60.4\%) \\ 
  $<$10 & 843 (41.7\%)& 102 (42.0\%) & 326 (42.4\%) & 234 (42.4\%) & 181 (39.6\%) \\ 
 \midrule
     \multicolumn{6}{l}{\textbf{Sex}} \\ 
     \hline
     Female & 968 (47.9\%)  & 103 (42.4\%) & 375 (48.8\%) & 268 (48.6\%) & 222 (48.6\%) \\ 
  Male & 1053 (52.1\%) &140 (57.6\%) & 394 (51.2\%) & 284 (51.4\%) & 235 (51.4\%) \\ 
   \bottomrule
   \label{t:demographic}
\end{tabular}
}
\end{table}

\subsection{Sensitivity Analysis and Model Validation}
First, we performed a sensitivity analysis on the number of estimated subgroups to the hyperparameter specification. We selected various smoothing parameter values $b$ to assess the sensitivity of our LASIR methods to kernel hyperparameters. Specifically, we considered $b\in \{1250, 300, 200, 120, 80\}$, which correspond to a range of $\rho$ values, i.e., $\rho \in\{0.02,0.04,0.05,0.06,0.08\}$ when the modified kernel is represented as $\tilde{\kappa} = \exp\{-a(\|v_1\|^2 + \|v_2\|^2) - ||v_1 - v_2||_2^2/2\rho^2\}$. A larger value of $b$ indicates a less smooth GP. To ensure consistency, we chose $L = 680$ as the number of basis functions such that the total variance explained by the basis function approximation is comparable to the kernel hyperparameter setting used in our real data analysis, i.e., $(h = 14, b = 200)$. Table~\ref{tab:sensitivity-subgroups} gives the number of individuals detected in each subgroup using different hyperparameter settings of smoothing parameter $b$, where the resulting subgroups have large overlaps. LASIR consistently identifies four subgroups across different hyperparameter combinations, with the number of individuals detected remaining generally consistent. The results demonstrate the insensitivity of subgroup detection to smoothing hyperparameter $b$ in our LASIR method.

\begin{table}[tp]
\centering
\caption{Number of individuals detected in each subgroup using different hyperparameter settings of smoothing parameter $b$. These results correspond to different correlation values between neighboring voxels associated with each value of $b$. The variance contribution rate is the proportion of explained variance of the reference ($a=0.01, b=200, L' = 1140$)}.
\begin{tabular}{|c|c|c|c|c|c|}
\hline
Smoothing parameter $b$ & $80$& $120$ & $200$ & $300$  & $1250$\\ \hline
Neighboring correlation & $0.98$& $0.97$&$0.96$ & $0.94$& $0.78$\\
\hline
Variance contribution rate & 61.8\% & 61.4\% & 61.0\% & 60.7\% & 60.2\%\\ \hline
Subgroup 1  & 265 & 270 & 243 & 340 & 226\\ \hline
Subgroup 2 & 832 & 781 & 769 & 717 & 706\\ \hline
Subgroup 3 & 455 & 501 & 552 & 449 & 587\\ \hline
Subgroup 4 & 469 & 469 & 457 & 515 & 502\\ \hline
\end{tabular}
\label{tab:sensitivity-subgroups}
\end{table}
Second, we validated the subgroup identification with a training-validating procedure. We started by using LASIR to estimate the subgroup labels $\{(\hat{\delta}_{ik})_{k=1}^K\}$ and stratifying individuals, and then repeatedly draw 50 subsamples from the entire dataset as training data, with each set containing 1,817 individuals, and reserved the remaining 104 individuals as a validation set. 

We compared projected prediction MSEs in the leave-out validation set given by three different models fitted to the training data. 1) Within-subgroup projection: given subgroup labels $\{(\hat{\delta}_{ik})_{k=1}^K\}$ estimated from the LASIR, the SVCM was fitted within each subgroup $k$, for $k = 1,\ldots, K$. 2) Without-subgroup projection: the SVCM was fitted to the training data without subgroups. And 3) Shuffled-subgroup projection: given LASIR subgroup labels, we shuffled the labels across subgroups, and the SVCM was fitted within each subgroup $k$.

\begin{figure}[h]
    \centering
    \includegraphics[scale=0.3,width=0.95\textwidth]{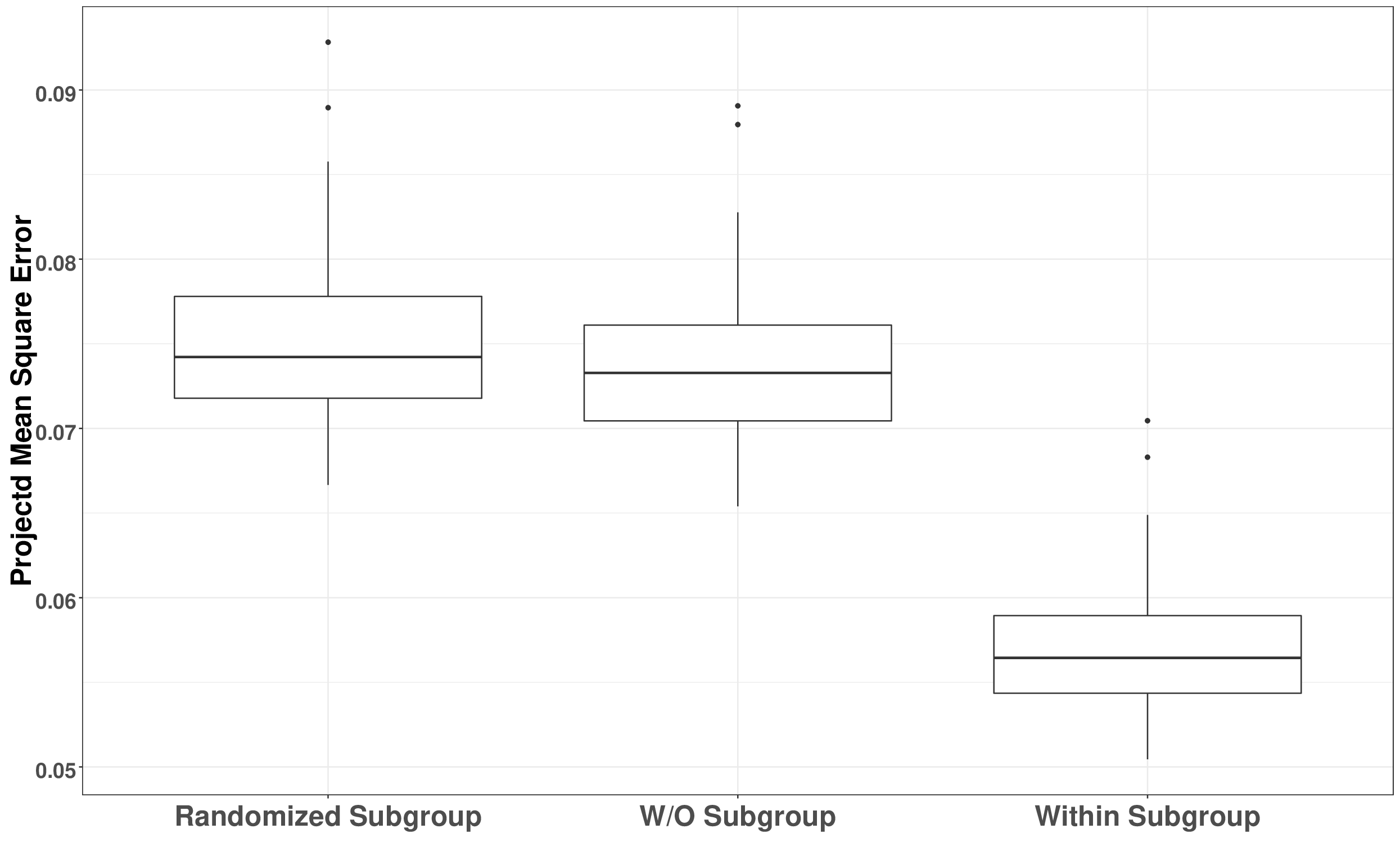}
    \caption{Boxplots of projected prediction mean squared error (MSE) over 50 data subsets as the leave-out validation, given by three approaches: 1) within-subgroup projection (right). 2) without-subgroup SVCM (middle); and 3) randomly shuffled (left).}
    \label{f:mse_cv}
\end{figure}

Figure \ref{f:mse_cv} shows the projected MSEs of different training-validating approaches. The within-subgroup validation provides the smallest projected MSE, followed by the without-subgroup validation approach. The projected MSE given by the shuffled-subgroup validation approach is slightly worse than the projected MSE given by the without-subgroup validation approach. This result validates the homogeneity of association within subgroups identified by LASIR. If the LASIR identifies the true latent subgroups, the within-subgroup projection should provide the smallest MSE. The shuffled-subgroup projection assigns incorrect subgroup labels to individuals and thus yields large MSEs. The without-subgroup projection is similar to one SVCM and can result in large values of projected MSE due to subgroup heterogeneity.

\section{Discussion}
\label{s:discuss}

We present a novel subgroup detection method for imaging data, LASIR, to estimate spatially varying coefficient functions of the group-specific coefficients in image-on-scalar regressions. Our LASIR method can capture both within-subgroup homogeneity and between-subgroup heterogeneity in functional brain activities and their associations with scalar characteristics. The proposed basis expansion approach and the SEM algorithm are computationally flexible and efficient for estimating the SVCs of interest. In our simulation experiments, LASIR can outperform existing methods for effect size estimation and subgroup identification in image-on-scalar regression models. From the analysis of fMRI data in the ABCD, the LASIR results offer new insights into the heterogeneous associations between functional brain activities and cognition measures among adolescents. Although several literature studies have focused on detecting the explicit demographic difference, e.g., across age and sex groups, in subjects' brain-wide activities associations \citep{assari2021association}, our LASIR method is a hierarchical model-based approach that automatically detects latent subgroup differences in both brain activities and their associations with subjects' cognitive, social or clinical characteristics. The latent subgroup method improves dimension reduction in brain-wide association studies without adding numerous interactions between characteristics and imaging measures into the model.

The identified subgroups show that heterogeneity exists in brain activities and their associations with cognitive scores. Consistent with the literature studies, the associations between cognitive ability and brain activities in specific brain functional networks, such as executive/reward/default mode networks, are heterogeneous across a large population~\citep{Marek2022}. The identified subgroups show distinct sociodemographic profiles, including race, income level, parental education level, and marital status, and indicate a potential disparity in the population of ABCD study subjects. The heterogeneity in such associations can also be moderated by sociodemographic risks such as social marginalization, racial profiling, and social discrimination. The identified population heterogeneity across sociodemographic subgroups may provide insights to balance the sample decomposition between the ABCD study with the targeted U.S. children population and achieve population generalizability.

There are several future directions of interest to improve LASIR. First, in the basis expansion approach, LASIR requires pre-determine some hyperparameter values, including the number of basis functions and the number of latent groups. There are demands to develop data-driven methods that automatically estimate hyperparameters. Second, the proposed model inference procedure based on asymptotics may not be valid for high-resolution images with limited sample sizes. Third, it would be very useful to develop LASIR under a fully Bayesian inference framework. However, posterior computation with large-scale imaging data needs new development. Variational inference methods can be a promising solution. Lastly, LASIR focuses on the heterogeneity in brain activation patterns and association studies and can be extended to account for other sources, such as variances. We could introduce an additional spatially-varying variance parameter for the random error $\epsilon_i(v)$, but the parameter estimation procedure can be challenging and requires further investigation in future research.




\begin{acks}[Acknowledgments]
We appreciate the assistance of Drs. Chandra S. Sripada and Mike Angstadt for the ABCD study data processing and sharing.
\end{acks}
\begin{funding}
This work was partially supported by the NIH grants R21HD105204 (Si, Lin, Kang), R01
DA048993 (Kang), R01 GM124061 (Kang),
and R01 MH105561 (Kang)
\vspace*{-8pt}
\end{funding}



\begin{supplement}

\section*{Supplement to ``Latent subgroup identification in Image-on-scalar regression"} The supplementary materials are available online, including the R code for data analysis and related R package. In the supplementary material, we provide supplemental information about the Hermite polynomials and basis function construction, the process of using the \texttt{GPfit} package to estimate the smoothing parameter, sensitivity analysis with varying hyperparameter values, and additional figures describing the detailed simulation and application results.
\end{supplement}

\bibliographystyle{imsart-nameyear} 
\bibliography{b1.bib}       

\end{document}